\documentclass[pra,aps,a4paper,notitlepage,twocolumn,longbibliography]{revtex4-1}

\usepackage{amssymb}
\usepackage{amsmath}
\usepackage{amsfonts}
\usepackage{graphicx}
\usepackage{bm}
\usepackage[dvipsnames]{xcolor}
\usepackage{natbib}
\usepackage{hyperref}
\usepackage{esint}

\newcommand{\corr}[1]{\langle #1\rangle}

\def\be{\begin{equation}}
\def\ee{\end{equation}}

\def\bs{\mathbf{s}}
\def\bt{\mathbf{t}}
\def\br{\mathbf{r}}

\def\bp{\mathbf{p}}

\def\bk{\mathbf{k}}
\def\bv{\mathbf{v}}
\def\bn{\mathbf{n}}

\def\bquasi{\mathbf{k}}
\def\quasi{k}
\def\bphon{\mathbf{q}}
\def\phon{q}

\begin{document}

\title{Electron-phonon relaxation in a model of a granular film}

\author{Nikolai A. Stepanov}
\affiliation{L.\ D.\ Landau Institute for Theoretical Physics, Chernogolovka 142432, Russia}

\author{Mikhail A.\ Skvortsov}
\affiliation{L.\ D.\ Landau Institute for Theoretical Physics, Chernogolovka 142432, Russia}

\date{\today}

\begin{abstract}
We study the electron-phonon relaxation in the model of a granular metal film, where the grains are formed by regularly arranged potential barriers of arbitrary transparency. The relaxation rate of Debye acoustic phonons is calculated, taking into account two mechanisms of electron-phonon scattering: the standard Fr\"ohlich interaction of the lattice deformation with the electron density and the interaction mediated by the displacement of grain boundaries dragged by the lattice vibration. At the lowest temperatures, the electron-phonon cooling power follows the power-law temperature dependence typical for clean systems but with the prefactor growing as the transparency of the grain boundaries decreases.
\end{abstract}

\maketitle

\section{Introduction}

Electron-phonon cooling significantly reduced at low temperatures has a pronounced impact on the performance metrics of various nanodevices, such as superconducting single photon detectors \cite{Goltsman,Semenov2021,Vodolazov}, hot electron bolometers \cite{Karasik2003,Tretyakov2011}, quantum phase slip devices \cite{Mooij,Astafiev,Astafiev-comment}, etc.

Quite generally, the heat flux transferred from 
electrons to phonons per unit volume can be written as \cite{Wellstood}
\be
\label{T^p-T^p}
  P = \Sigma (T^p_\text{el}-T^p_\text{ph}) ,
\ee
where the temperature exponent $p$ is determined by the effective dimensionality $d$ of the sample and and the degree of disorder. In the clean case, $p=d+2$ (since the phase volume of thermal phonons $\propto T^d$, the number of involved electrons $\propto T$, and a typically transmitted energy in the collision $\propto T$) 
\cite{Wellstood,Hekking,Viljas,Cojocaru}. 
Generalization of this result to the case of \emph{homogeneously disordered} metals has been intensively studied in literature \cite{Pippard,Schmid,Tsuneto,ReyzerSergeev,YudsonKravtsov,Shtyk,FeigelmanKravtsov}.
The presence of point-like disorder modifies the electron-phonon relaxation rate at low temperatures when $\phon_Tl\ll1$, where $\phon_T=T/c$ is the thermal phonon wave vector, $c$ is the sound velocity, and $l$ is the electron elastic mean free time. In this regime, as shown by Pippard \cite{Pippard}, motion of impurities dragged by lattice vibrations results in the suppression of the relaxation rate by the factor of $\phon_Tl\ll1$, reducing the temperature exponent $p\to p+1$ \cite{Schmid,Tsuneto}. On the other hand, if impurities are not completely dragged by the lattice, the electron-phonon coupling could be enhanced with the exponent $p\to p-1$ \cite{Sergeev}, though such a model can hardly be justified microscopically. Note, however, that although the condition $\phon_Tl\ll1$ is satisfied for most metals at low temperatures, the heat transfer between electrons and phonons is often well described by the exponent $p=5$ corresponding to the clean three-dimensional (3D) situation \cite{Pekola-review}.

The integral quantity $\Sigma T^p$ can be conveniently expressed in terms of the \emph{phonon} relaxation rate (ultrasound attenuation rate) $\tau_\text{ph}^{-1}(\omega)$, which under rather general assumptions is temperature-independent \cite{Shtyk}, by the relation
\be
\label{W-ph}
  \Sigma T^p
  =
  \int_0^\infty d\omega\, \omega \, \nu_\text{ph}(\omega) 
  \tau_\text{ph}^{-1}(\omega) N(\omega,T) ,
\ee
where $N(\omega,T)=1/(e^{\omega/T}-1)$ is the Bose-Einstein distribution function and $\nu_\text{ph}(\omega)=s_d\omega^ {d-1}/(2\pi c)^d$ is the phonon density of states in $d$ dimensions (with $s_d$ being the area of the unit sphere).
An alternative description widely used in literature operates with the \emph{electron} scattering rate on lattice vibrations, $\tau_\text{e-ph}^{-1}(E,T)$, which is a complicated function of the electron energy and temperature.
Equation \eqref{T^p-T^p} corresponds to a power-law dependence $\tau_\text{ph}^{-1}(\omega)=\alpha\omega^{p-d-1}$ with the heat transfer coefficient
\be
\label{Sigma}
  \Sigma
  =
  s_d\Gamma(p)\zeta(p) \alpha/(2\pi c)^d 
\ee
and the electron scattering rate at the Fermi energy given by $\tau_\text{e-ph}^{-1}(T) \equiv \tau_\text{e-ph}^{-1}(0,T) = \eta\Sigma T^{p-2}/\nu_0$, where $\eta=(2-2^{3-p})\Gamma(p-2)\zeta(p-2)/\Gamma(p)\zeta(p)$ and $\nu_0$ is the density of states (DOS) at the Fermi level \cite{Wellstood,Rammer}. Here $\Gamma$ and $\zeta$ are the gamma and Riemann zeta functions, respectively.

Recently, electron-phonon interaction has been actively studied in strongly disordered NbN films \cite{Sidorova,Dane,Lomakin} used in photon detectors. It was found that for $T>10\,\text{K}$, when phonons are effectively three-dimensional (3D), the scattering rate is well described by $\tau_\text{e-ph}^{-1}(T)=\gamma T^3$ with disorder-independent parameter $\gamma$ by an order of magnitude exceeding its value in a clean system \cite{Lomakin}. Such a behavior cannot not described by existing theories and it was suggested that it is mediated by the electron-phonon interaction at the crystallite boundaries.

Motivated by these experimental results, we aim at incorporating scattering on grain boundaries into the theory of electron-phonon interaction. For this purpose we will consider the simplest case of regularly placed granules and calculate the phonon relaxation rate $\tau_\text{ph}^{-1}(\omega)$ for arbitrary transparency of boundaries between them.

The paper is organized as follows. In Sec.\ \ref{S:Model} we introduce the model and obtain the basic ingredients required for calculating the relaxation rate. Various simplifications arising in the low-temperature limit are discussed in Sec. \ref{S:Low-temp} and then used in Sec.\ \ref{S:vs-kappa} to obtain the dependence of $\tau_\text{ph}(\omega)$ on the barrier transparency, both analytically and numerically, 
and consider the effect of inelastic broadening of the electron spectrum.
The results are summarized and discussed in Sec.\ \ref{S:Conclusion}.

\section{Model and basic relations}
\label{S:Model}

\subsection{Model}

Electron-phonon relaxation in a granular medium will be modelled by the Hamiltonian
\begin{equation}
\label{H-general}
H = {\cal{H}}_\text{el} + {\cal{H}}_\text{ph} + {\cal{V}} ,
\end{equation}
describing the electron (${\cal{H}}_\text{el}$) and phonon (${\cal{H}}_\text{ph}$) subsystems, and the interaction between them (${\cal{V}}$).

The single-electron Hamiltonian
\be
\label{H-el-1}
{\cal{H}}_\text{el}
=
\int d^2r \, \psi^\dagger 
\left[
\bp^2/2m
+
U(x)+U(y)
\right]
\psi
\ee
describes motion in the field of $\delta$-barriers regularly placed along the $x$ and $y$ axes:
\be
\label{H-el-2}
U(x)=
\frac{\varkappa}{m}
\sum_n\delta(x-na) .
\ee
The parameter $\varkappa$, with the dimension of momentum, determines the tunnel transparency of the boundary at normal incidence with momentum $p$:
\be
\label{T(p)}
  {\cal T}(p) = \frac{p^2}{p^2+\varkappa^2} .
\ee
The period of the square lattice of the barriers (granule size) is assumed to be much larger than the Fermi wavelength, $a\gg\lambda_F$.

The Hamiltonian of acoustic phonons written in terms of the displacement $u_\alpha(\br)$ and momentum $\pi_\alpha(\br)$ fields has the form \cite{LL7}
\begin{equation}
\label{H-ph}
{\cal{H}}_\text{ph}=
\frac{\rho}{2}\int d^2r 
\left[
\pi_\alpha^2
+(c^2_\text{\sc{l}}-c^2_\text{\sc{t}})(\partial_\alpha u_\alpha)^2
+c^2_\text{\sc{t}}\left(\partial_\alpha u_\beta\right)^2
\right],
\end{equation}
where $\rho$ is the crystal density and $c_\text{\sc{l},\sc{t}}$ are the longitudinal and transverse sound velocity. 
In the two-dimensional (2D) case considered, the standard quantization of the Hamiltonian \eqref{H-ph} yields two independent acoustic phonon modes with the linear dispersion $\omega_\nu(\bphon)=c_\nu \phon$ and polarization vectors $\bm{\epsilon }^{(\nu)}$ pointing, respectively, along and perpendicular to the phonon propagation direction $\bphon$.

In the considered model, there are two mechanisms of the electron-phonon interaction. The first one is the standard Fr\"ohlich interaction of the local electron density with the divergence of the longitudinal displacement of the lattice \cite{Frohlich,Gantmacher}:
\begin{equation}
\label{e-ph-1}
{\cal{V}}_{1}
=
\zeta E_\text{F}
\int d^2r \,
\psi^\dagger
(\partial_\alpha u_\alpha)
\psi ,
\end{equation}
where $E_\text{F}$ is the Fermi energy and $\zeta= 2/d$ is the dimensionless parameter of the electron-phonon interaction calculated in the RPA approximation \cite{Gantmacher,Ziman,Rammer}.
The second mechanism is the displacement of grain boundaries due to lattice vibrations that amounts to replacing the coordinate $r_\alpha-na$ in Eq.\ \eqref{H-el-2} by $r_\alpha-na-u_\alpha(\br)$. 
In the linear order in the phonon field, the corresponding electron-phonon interaction reads
\begin{equation}
\label{e-ph-2}
{\cal{V}}_{2}
=
\frac{\varkappa}{m} 
\int d^2r\,\psi^\dagger\left(\sum_{n,\alpha}
 \delta^\prime(r_\alpha-na)
u_\alpha(\br)
 \right)\psi ,
\end{equation}
where the index $\alpha$ takes two values: $x$ and $y$.

The second type of the electron-phonon interaction, ${\cal V}_2$, inevitably arises in disordered metals, where electrons are subject to an external potential produced by vacancies or crystalline defects coupled to the lattice and dragged by the motion of the latter \cite{Schmid,ReyzerSergeev,YudsonKravtsov,FeigelmanKravtsov}.

\subsection{Electronic states}
\label{SS:electronic_states}

\begin{figure}
   \centering
\includegraphics[width=0.8\linewidth]{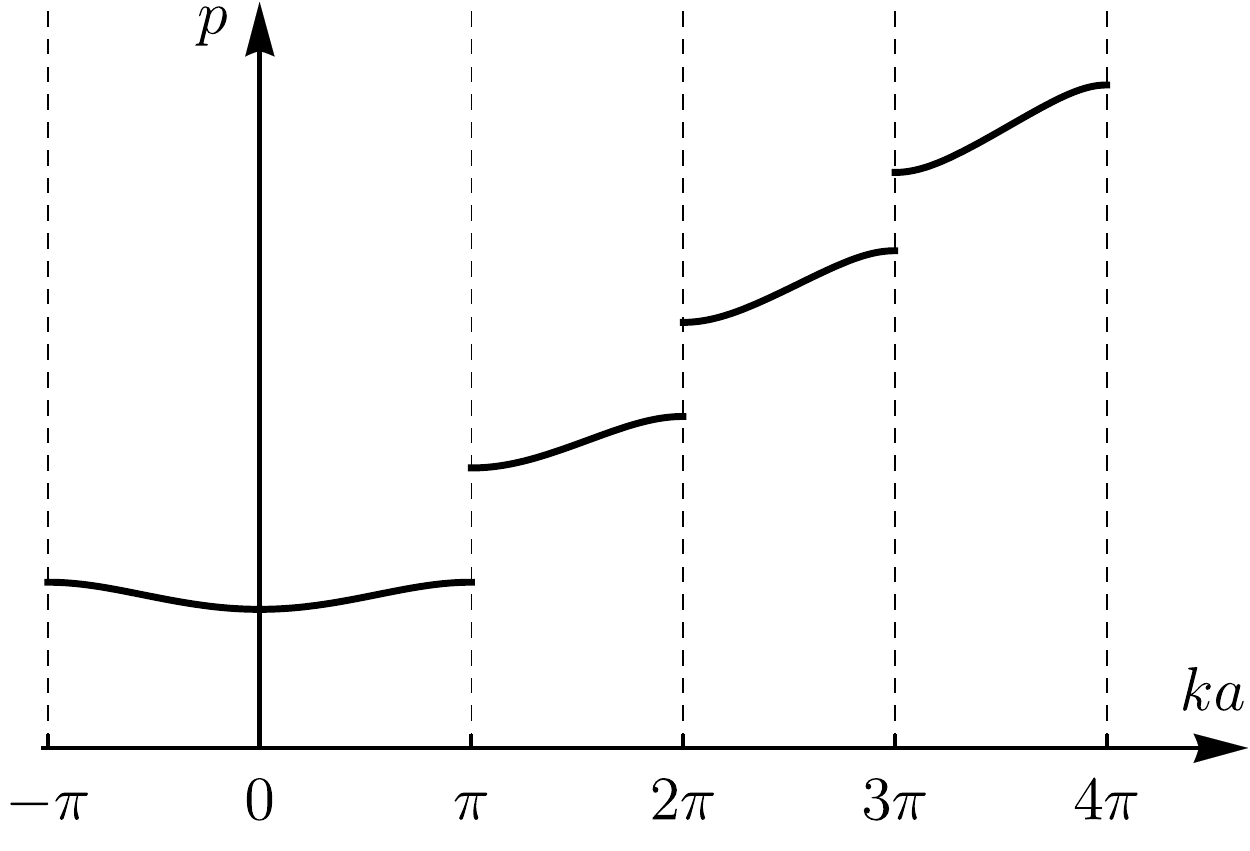}
    \caption{The momentum $p$ vs the quasimomentum $k$ in the extended zone scheme for the 1D Dirac comb potential, according to Eq.\ \eqref{q-p relation}, for $\varkappa a=10$.
    }
    \label{F:p(q)}
\end{figure}

The chosen microscopic model of electrons in a periodic granular medium governed by the Hamiltonian~\eqref{H-el-1} allows for the separation of variables: For each of the axes, the problem reduces to the well-known one-dimensional (1D) Dirac comb problem \cite{KronigPenney}. Its properties relevant for the calculation of the electron-phonon relaxation rate are summarized below.

Electron motion in a 1D periodic potential is described by the Bloch quasimomentum $\quasi$. We will use the extended zone scheme \cite{Kittel} when the quasimomentum is not limited to the first Brillouin zone $(-\pi/a,\pi/a)$, but takes all real values.
In the coordinate representation, the wave function $|\quasi\rangle$ is piecewise smooth, given on the interval $x_n<x<x_{n+1}$ by the superposition of two plane waves:
\be
\label{wf}
  \langle x|\quasi\rangle
  =
  C_\quasi e^{i\quasi x_{n+1/2}} 
  \sum_{\sigma=\pm1} 
  \sin\frac{(p + \sigma \quasi)a}2
    e^{i\sigma p(x-x_{n+1/2})}
  ,
\ee
where $x_n=na$ is the position of the $n$-th wall, while $x_{n+1/2}$ corresponds to the midpoint between the two walls.
The relation between the momentum $p$ and the quasimomentum $\quasi$ has the form
\be
\label{q-p relation}
\cos \quasi a=\cos pa+\varkappa\frac{\sin pa}{p}.
\ee
When solving the transcendental equation (\ref{q-p relation}) on $p(\quasi)$ in the extended zone scheme, $p$ and $\quasi$ must belong to the same Brillouin zone, see Fig.\ \ref{F:p(q)}.
The dispersion relation reads $E(\quasi)=p^2(\quasi)/2m$. 
For the wave functions normalized to the quasimomentum, $\left<\quasi|\quasi'\right>=2\pi\delta\left(\quasi-\quasi'\right)$, the real coefficient $C_\quasi$ in Eq.\ \eqref{wf} reads [here $p=p(\quasi)$]
\begin{equation}
\label{C-def}
  \frac{1}{C_\quasi^2}
  =
  1-\cos \quasi a\cos pa
  + \varkappa a\left(\frac{\sin pa}{pa}\right)^{2}
  .
\end{equation}

The wave functions of the 2D Hamiltonian \eqref{H-el-1} 
are given by $|\bquasi\rangle = |\quasi_x\rangle|\quasi_y\rangle$, with the corresponding energy $E(\bquasi)=[p^2(\quasi_x)+p^2(\quasi_y)]/2m$.

The spectrum in the 1D Dirac comb potential is organized in a sequence of bands separated by gaps, with the DOS possessing van Hove singularities at the gap edges. Each band is formed by the states with the quasimomentum from the corresponding Brillouin zone.

\begin{figure}
\centering
\includegraphics[width=0.99\columnwidth]{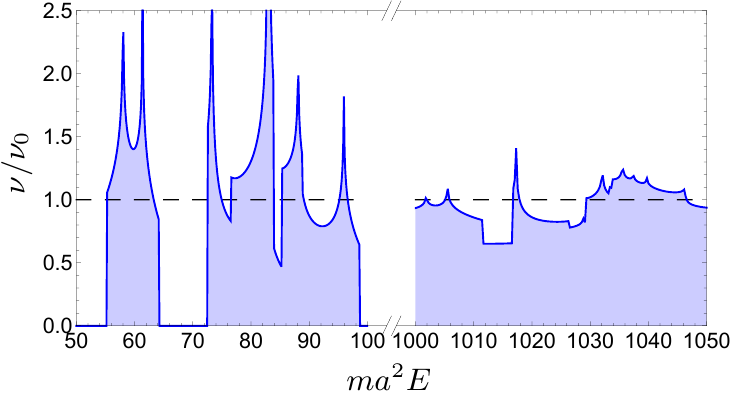}
    \caption{
    Electron density of states (in units of $\nu_0=m/2\pi$) for the 2D model \eqref{H-el-1} of regularly placed $\delta$-barriers with $\varkappa a=25$.}
    \label{fig:DOS}
\end{figure}

In the 2D model considered, the states of a 2D Brillouin zone (a square of size $\pi/a\times\pi/a$ in the quasimomentum space), which will be refereed to as a \emph{plaquette}, also form a spectral band with logarithmic van Hove singularities in the middle of the band and regular behavior at the edges. But now different spectral bands are allowed to overlap that produces a rather complicated DOS structure with a large number of van Hove singularities, see Fig.\ \ref{fig:DOS}.
In the vicinity of a given energy $E$ the spectrum may be either gapped or gapless depending on the strength of the $\delta$-function barrier. 
Which of the two regimes is realized depends on the value of the dimensionless parameter 
\be
\label{gamma-def}
  \gamma=(p/\varkappa)(pa), 
\ee
where the first factor $p/\varkappa$ is the fraction of finite-DOS regions for $\varkappa\gg p$, while the second factor $pa\gg1$ estimates the number of plaquettes at an energy $E=p^2/2m$.

The \emph{good-metal} regime corresponds to strong coupling between the grains, $\gamma\gg1$ (right part of Fig.\ \ref{fig:DOS}). In this case, at a given energy near $E_\text{F}$ many bands do contribute to the DOS, which therefore can be viewed as a constant slightly perturbed by quasiregularly placed van Hove singularities. It is this regime we are going to consider below.

In the opposite case of weak coupling, $\gamma\lesssim1$ (left part of Fig.\ \ref{fig:DOS}), the spectrum becomes gapped and the system may demonstrate either metallic or insulating properties depending on the position of the Fermi energy and temperature.

\subsection{Fermi's golden rule}

\begin{figure}
\centering
\includegraphics[width=0.7\columnwidth]{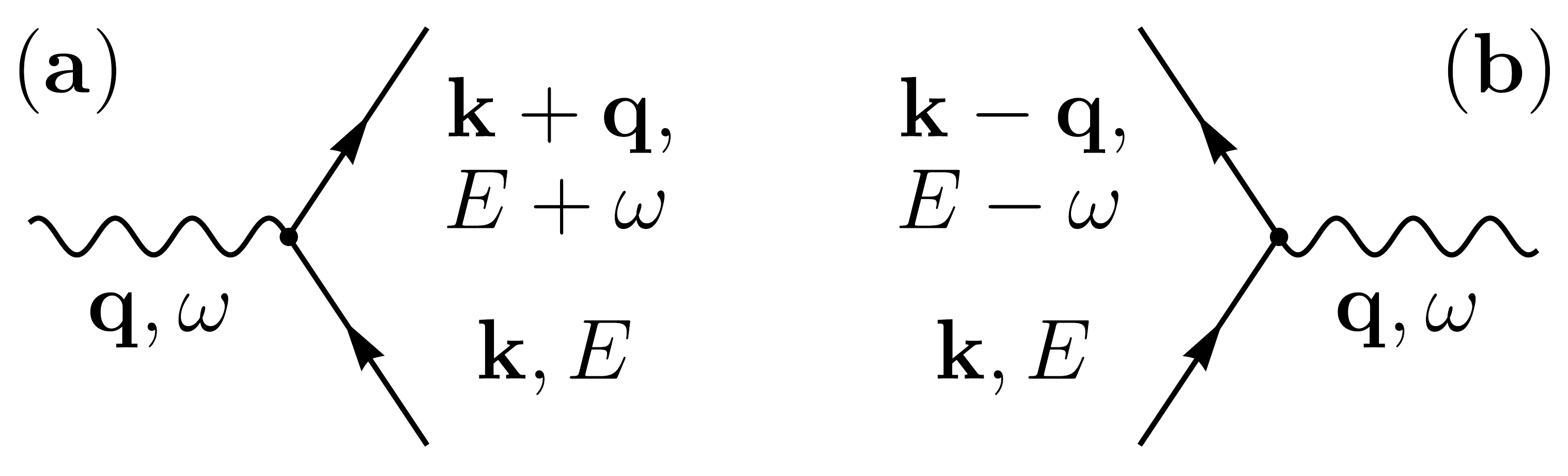}
    \caption{The processes of phonon absorption (a) and emission (b) by an electron-hole pair.}
    \label{fig:diagramm}
\end{figure}

The basic processes of electron-phonon interaction are (i) the decay of a phonon with momentum $\bphon$ and polarization $\nu$, which generates an electron-hole pair with momenta $\bquasi_2$ and $\bquasi_1$ (Fig.\ \ref{fig:diagramm}a), and (ii) a dual process of phonon emission (Fig.\ \ref{fig:diagramm}b).
The rates of such processes due to the interaction ${\cal{V}}_i$ can be found by Fermi's golden rule:
\begin{multline}
\label{GR-CF}
w_{\nu,i}^{(s)}(\bphon)
=
2\pi \int (d\bquasi_1)(d\bquasi_2)
|M_{\nu,i}|^2
\delta[E_2-E_1-\omega_\nu(\bphon)]
\\{}
\times
(2\pi)^2\hat\delta(\bphon+\bquasi_1-\bquasi_2)
  \Phi_s(E_1,E_2) .
\end{multline}
Here $s=\text{out}$ (in) corresponds to the processes of phonon decay (emission) characterized by different combinations of the Fermi-Dirac distribution function $f(E)$: 
$\Phi_\text{out}(E_1,E_2) = f(E_1)[1-f(E_2)]$
and
$\Phi_\text{in}(E_1,E_2) = [1-f(E_1)]f(E_2)$.
Kinetic coefficients are determined by the difference of out- and in-contributions. The phonon relaxation (attenuation of ultrasound) rate is given by \cite{Wellstood}
\be
\label{GR-CF-relax}
  w_{\nu,i}(\bphon) 
  = 
  w_{\nu,i}^\text{(out)}(\bphon) - w_{\nu,i}^\text{(in)}(\bphon) ,
\ee
which corresponds to $\Phi(E_1,E_2) = f(E_1)-f(E_2)$. Averaging $w_{\nu,i}(\bphon)$ at the angles, we get $\tau_\text{ph}^{-1}(\omega=c_\nu \phon)$.

In Eq.\ \eqref{GR-CF},
$\hat\delta(\bphon)=\hat\delta(\phon_x)\hat\delta(\phon_x)$,
where we introduced a periodically-continued $\delta$ function ensuring momentum conservation in the presence of Umklapp processes,
\be
  \hat\delta(\phon)
  =
  \sum_m \delta(\phon-2\pi m/a) ,
\ee
while $M_{\nu,i}$ is the matrix element of the interaction ${\cal V}_i$ with the $\delta$ function excluded.

\subsection{Matrix elements}

Quantizing phonons in the standard way \cite{Gantmacher}, we find
\be
\label{M-def}
(2\pi)^2\hat\delta(\bphon+\bquasi_1-\bquasi_2)
M_{\nu,i}
=
\sum_\alpha 
\frac{\epsilon^{(\nu)}_\alpha\corr{\bquasi_{2}|V_i^\alpha(\br) e^{i\bphon\br}|\bquasi_1}}
{\sqrt{2\rho\omega_\nu(\bphon)}}
.
\ee
The functions $V_i^\alpha(\br)$ corresponding to the interactions \eqref{e-ph-1} and \eqref{e-ph-2} have the form
\be
V^\alpha_1 = \zeta E_\text{F} \phon_\alpha ,
\qquad
V^\alpha_2=
\frac{\varkappa}{m}\sum_n\delta^\prime(r_\alpha-na).
\ee

Since the wave functions $|\bquasi\rangle$ are factorized, to find $M_{\nu,i}$ from Eq.\ \eqref{M-def}, it is sufficient to compute the matrix elements of 1D operators
\be
  \Pi_1 = e^{i\phon x} ,
\qquad
  \Pi_2 = i \sum_n \delta'(x-na) e^{i\phon x} 
\ee
betwen the states \eqref{wf}. A simple calculation gives
\be
  \corr{\quasi_2|\Pi_i|\quasi_1}
  =
  2\pi\hat\delta(\phon+\quasi_1-\quasi_2) 
  C_{\quasi_1}C_{\quasi_2}
  J_i(\phon;\quasi_1,\quasi_2) ,
\ee
where $J_1(\phon;\quasi_1,\quasi_2)$ reads
\begin{multline}
  J_1(\phon;\quasi_1,\quasi_2)
  =
  \sum_{\sigma,\tau=\pm1}
  \sin\frac{(p_1 + \sigma \quasi_1)a}2
  \sin\frac{(p_2 + \tau \quasi_2)a}2
\\{}
  \times
  \frac{\sin[(\sigma p_1-\tau p_2-\quasi_1+\quasi_2)a/2]}{(\sigma p_1-\tau p_2+\phon)a/2} ,
\end{multline}
and $J_2(\phon;\quasi_1,\quasi_2)$ has the form
\begin{multline}
  a J_2(\phon;\quasi_1,\quasi_2)
  =
  p_1 \sin p_2a \sin \quasi_1a
   - p_2 \sin p_1a \sin \quasi_2a
\\{}
  + \phon \sin p_1a \sin p_2a .
\end{multline}

In terms of $J_i^\alpha=J_i(\phon_\alpha;\quasi_{1\alpha},\quasi_{2\alpha})$, the matrix elements $M_{\nu,i}$ take the following form:
\begin{subequations}
\label{M-gen}
\begin{gather}
\label{M1-gen}
M_{\text{\sc{l}},1}
=
\frac{\zeta E_\text{F} \phon\, {\cal C} J_1^xJ_1^y}
{\sqrt{2\rho c_\text{\sc{l}} \phon}}
,
\qquad
M_{\text{\sc{t}},1} = 0 ,
\\
\label{M2-gen}
M_{\nu,2}
=
\frac{-i
(\varkappa/m) \, {\cal C} \bigl[\epsilon^{(\nu)}_xJ_2^xJ_1^y+\epsilon^{(\nu)}_yJ_2^yJ_1^x\bigr]}{\sqrt{2\rho c_\nu \phon}}
,
\end{gather}
\end{subequations}
where ${\cal C} = C_{\quasi_{1x}} C_{\quasi_{1y}} C_{\quasi_{2x}} C_{\quasi_{2y}}$.

Note the phase difference of $\pi/2$ between the matrix elements for the interactions ${\cal V}_1$ and ${\cal V}_2$. This is the reason for vanishing of the cross term proportional to $M_{\nu,1}^*M_{\nu,2}+M_{\nu,1}M_{\nu,2}^*$.
Otherwise it would be impossible to separate the contributions from the vertices ${\cal V}_1$ and ${\cal V}_2$, and an analog of Eq.\ \eqref{GR-CF} would have to contain $|M_{\nu,1}+M_{\nu,2}|^2$ for the net relaxation rate.

\section{Ultrasound attenuation rate at~low temperatures}
\label{S:Low-temp}

\subsection{Temperature regimes}

The main contribution to the energy transfer between electron and phonon systems is provided by thermal phonons with frequency $\omega\sim T$ and momentum $\phon_T\sim T/c$. Therefore, the physics of electron-phonon relaxation is determined by the relation between the wavelength of the thermal phonon $\phon_T^{-1}$ and the granule size $a$. At high temperatures, $T\gg c/a$, the interaction actually occurs inside a single granule. In the extended zone scheme, that means that the quasimomenta of the electron and the hole,  $\bquasi_1$ and $\bquasi_2$, belong to different, well-separated Brillouin zones.
On the other hand, at low temperatures, 
\be
\label{low-T}
  T\ll c/a, 
\ee
the wavelength of the thermal phonon is large, and the elementary act of scattering involves electrons from different granules. In this regime, $\bquasi_1$ and $\bquasi_2$ belong to the same Brillouin zone that considerably simplifies the calculation, allowing one to neglect the Umklapp processes.

\subsection{Ultrasound attenuation rate at low temperatures}

To calculate the phonon relaxation rate from the general formulae \eqref{GR-CF} and \eqref{GR-CF-relax} in the limit $T\ll c/a$, one should simplify the matrix elements \eqref{M-gen}, as well as understand the kinematics of the process determined by the energy and momentum conservation laws.
In the limit of a small phonon momentum, $\phon\to0$, the electron and hole quasimomenta almost coincide,
$\bquasi_1\approx\bquasi_2\approx\bquasi$, and we can limit ourselves to the leading asymptotics of the functions ${\cal C}$ and $J$: 
$\,{\cal C} = C_{\quasi_x}^2 C_{\quasi_y}^2$,
$J_1^\alpha=C^{-2}_{\quasi_\alpha}$, 
$J_2^\alpha=i \phon_\alpha \beta(\quasi_\alpha) C^{-2}_{\quasi_\alpha}$,
where we introduced the function
\be
\label{beta-def}
  \beta(\quasi) 
  = 
  \varkappa \, \sin^2(pa)[1-\sin^2(pa)/(pa)^2] C^4_{\quasi}
  .
\ee
Then the matrix elements $|M_{\nu,i}(\bphon,\bquasi)|^2$ in Eqs.\ \eqref{M-gen} simplify to
\begin{subequations}
\label{matelem2}
\begin{gather}
|M_{\text{\sc{l}},1}|^2
=
(\zeta E_\text{F})^2 \phon / 2\rho c_\text{\sc{l}}
,
\\
\label{M-kappa-L2}
|M_{\text{\sc{l}},2}|^2
=
(\varkappa/m)^2
\bigl[n_x^2 \beta(\quasi_x) + n_y^2 \beta(\quasi_y) \bigr]^2 
\phon / 2\rho c_\text{\sc{l}},
\\
\label{M-kappa-T2}
|M_{\text{\sc{t}},2}|^2
=
(\varkappa/m)^2
n_x^2n_y^2\bigl[\beta(\quasi_x) - \beta(\quasi_y) \bigr]^2
\phon / 2\rho c_\text{\sc{t}} ,
\end{gather}
\end{subequations}
where $\bn=\bphon/\phon$ is the direction of phonon propagation.

\begin{figure}
\centering
\includegraphics[width=0.9\linewidth]{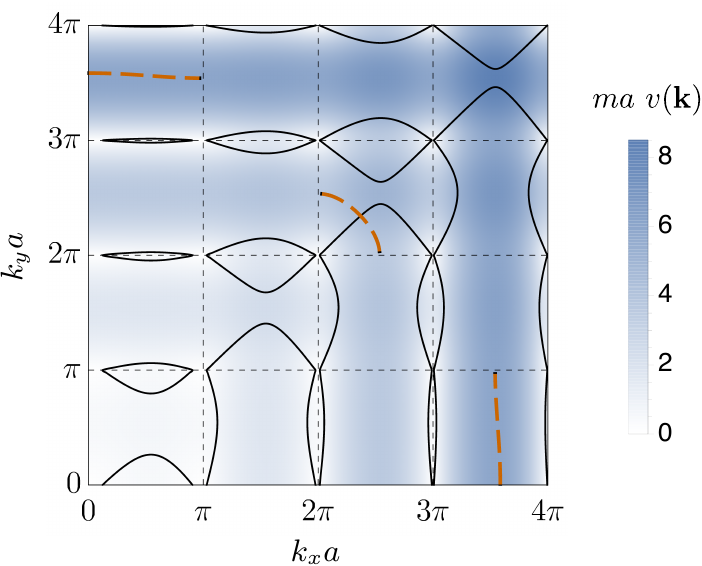}
\caption{Dependence of the velocity modulus $v(\bk)$ on $\bk$ (obtained for $\varkappa a=20$). The velocity clearly vanishes in the corners of the Brillouin zones (plaquettes). Shown by black is the line $\Gamma_\bn$ defined by Eq.\ \eqref{ECL-curve} with $c=0.1/ma$ and the vector $\bn$ making the angle $3\pi/4$ with the $x$ axis. Dashed orange line is the Fermi level $\Phi$ for $E_\text{F}=78/ma^2$.
}
\label{F:plaquettes}
\end{figure}

Solving the momentum conservation law by writing $\bquasi_{1,2}=\bquasi\mp\bphon/2$, we linearize the energy conservation law at small $\bphon$: 
$
\omega^{(\nu)}(\bphon)=E(\bquasi_2)-E(\bquasi_1)\approx \bphon \bv(\bquasi) ,
$
where $\bv(\bquasi)=\partial E(\bquasi)/\partial\bquasi$ is the electron velocity. It is more convenient to express the result in terms of the vector $\bn=\bphon/\phon$:
\be
\label{ECL-curve}
\bn\bv(\bquasi)=c_\nu .
\ee
In the $d$-dimensional case, Eq.\ \eqref{ECL-curve} defines a $(d-1)$-dimensional surface $\Gamma_{\bn}$ in the $\bquasi$ space, the shape of which is determined by the dispersion $E(\bquasi)$, see Fig.\ \ref{F:plaquettes}. For an isotropic parabolic spectrum, $\Gamma_{\bn}$ is a hyperplane orthogonal to the vector $\bn$. The main contribution to the phonon lifetime at low temperatures is determined by the intersection of the surface $\Gamma_{\bn}$ with the Fermi surface $\Phi$.

In the 2D case under consideration, the curves $\Gamma_{\bn}$ and $\Phi$ can intersect at some set of \emph{points}, with the number of them depending on the phonon propagation direction $\bn$. Consider the contribution of such an intersection, say, occurred at $\bquasi_0$, to the phonon decay rate. Let us parametrize the deviation from this point as
\be
\bquasi=\bquasi_0+\quasi_s \bs+\quasi_t \bt ,
\ee
with the unit vectors $\bs$ and $\bt$ being tangent and perpendicular to the curve $\Gamma_\bn$, respectively:
\be
\bt=
|\partial_\bquasi(\bn\bv)|^{-1} \,
\partial_\bquasi(\bn\bv),
\qquad 
\bs\bt=0.
\ee

Calculation of the contribution from the vicinity of the point $\bquasi_0$ to the phonon relaxation rate via Eqs.\ \eqref{GR-CF} and \eqref{GR-CF-relax} reduces to taking a 2D integral over $d\quasi_sd\quasi_t$. The transverse momentum $\quasi_t$ is pinned to zero by the delta-function term $\delta(\phon\quasi_t |\partial_\bquasi(\bn\bv)|)$.
In calculating the integral over the longitudinal momentum $\quasi_s$, we switch to integration over $dE=(\bv\bs)d\quasi_s$, which gives for $\Phi(E_1,E_2) = f(E_1)-f(E_2)$ the temperature-independent factor $\omega_{\nu}$.
As a result, we obtain for the contribution of the vicinity of $\bquasi_0$:
\be
\label{GR-q0}
  w_{\nu,i}(\bphon,\bquasi_0)
  =
  \frac{c_\nu|M_{\nu,i}(\bphon,\bquasi_0)|^2}
  {2\pi |[\bv,\partial_\bquasi(\bn\bv)]|}
  ,
\ee
where $\bv$ in the denominator is also taken in $\bquasi=\bquasi_0$.

To obtain the low-temperatures relaxation rate, one should determine all $\bquasi_0$ points and sum Eq.\ \eqref{GR-q0} over them:
\be
  w_{\nu,i}(\bphon)
  =
  \sum_{\bk_0} w_{\nu,i}(\bphon,\bquasi_0) .
\ee

\subsection{Averaging over angles}

The anisotropy of the considered model manifested in a complex structure of the Fermi surface $\Phi$ (see Figs.\ \ref{fig:DOS} and \ref{F:plaquettes}) makes the relaxation rate $w_{\nu,i}(\bphon)$ dependent not only on the absolute value of the phonon momentum $q$ but on its direction $\bn$ as well. Since the net rate of the electron-phonon heat flux involves integration over all directions, we will consider here the angle-averaged quantity
\be
  w_{\nu,i}(\phon)  
  =
  \corr{w_{\nu,i}(\phon\bn)}_\bn .
\ee

When the vector $\bn$ is rotated, the point $\bquasi_0$ moves along the Fermi line. Therefore, averaging over $\bn$ reduces to a properly weighted integration along the Fermi surface. To determine the appropriate measure, we take the derivative of Eq.\ \eqref{ECL-curve}:
\begin{equation}
\label{dd1}
  d\bn \,\bv + \bn (\partial\bv/\partial \quasi_l) d\quasi_l = 0 ,
\end{equation}
with $\quasi_l$ being the coordinate along the Fermi line. Writing $d\bn=[\hat z,\bn]d\theta$, where $\theta$ is the angle the vector $\bn$ makes with the $x$ axis, and $\hat z$ is a unit vector orthogonal to the plane, we transform the first term in Eq.\ \eqref{dd1} as $d\bn\,\bv = |[\bn,\bv]| d\theta = \sqrt{v^2-c_\nu^2} \, d\theta$. The scalar product in the second term, as one can readily see, is $\bn (\partial\bv/\partial \quasi_l) = | [\bv, \partial_\bquasi(\bn\bv)] |/v$. 
Hence, we find $d\theta/dk_l = | [\bv, \partial_\bquasi(\bn\bv)] |/v\sqrt{v^2-c_\nu^2}$.

Now we are in a position to average Eq.\ \eqref{GR-q0}, that is, to integrate over $d\theta/2\pi$. Switching to integration over $d\quasi_l$ we see that the factor of $|[\bv, \partial_\bquasi(\bn\bv)]|$ in $d\theta/dk_l$ cancels the same factor in the denominator of Eq.\ \eqref{GR-q0}, and we obtain for the angle-averaged relaxation rate:
\be
\label{GR-av-n}
  w_{\nu,i}(\phon)
  =
  \frac{c_\nu}{2\pi^2}
  \int_\text{FL}
  \frac{d\quasi_l \, 
  |M_{\nu,i}(\phon\bn(\bquasi),\bquasi)|^2 }
  {v(\bquasi)\sqrt{v^2(\bquasi)-c_\nu^2}}
  ,
\ee
where the integral is evaluated along the Fermi line, 
and the vector $\bn(\bquasi)$ is one of the two solutions of Eq.\ \eqref{ECL-curve} for a given $\bk$ [the second solution gives the same contribution that has already been taken into account in the prefactor of Eq.\ \eqref{GR-av-n}]. Regions of Fermi surface with $v(\bquasi)<c_\nu$, where the decay process is prohibited by the conservation laws, should be excluded from the integral \eqref{GR-av-n}.

\subsection{Averaging over the Fermi energy}

The angle-averaged relaxation rate defined by Eq.\ \eqref{GR-av-n} is a rather complicated function of the Fermi energy, qualitatively resembling the DOS dependence on $E_\text{F}$ (see discussion in Sec.\ \ref{SS:electronic_states}).
In the \emph{good-metal} regime realized at $\gamma\gg1$ [see Eq.\ \eqref{gamma-def}],
it contains a large smooth background and a small oscillating component due to quasiregularly placed van Hove singularities,
as shown in the right part of Fig.\ \ref{fig:DOS}.
These singularities arise due to periodicity of the model and will be smeared or completely destroyed by disorder or irregularity inevitably present in real samples.

To get rid of the (small) oscillating component, we average the angle-averaged relaxation rate $w_{\nu,i}(\phon)$ over the Fermi level:
\be
\label{gamma-1}
\corr{w_{\nu,i}(\phon)}_{E_\text{F}}
=
\int_{E_\text{F}-\Delta E_\text{F}/2}^{E_\text{F}+\Delta E_\text{F}/2}
\frac{dE_\text{F}}{\Delta E_\text{F}} w_{\nu,i}(\phon) .
\ee
The width of the distribution $\Delta E_\text{F}$ is chosen such that $\pi v_\text{F}/a\ll\Delta E_\text{F}\ll E_\text{F}$, where the first condition means that the width of the averaging area is large enough to include many plaquettes (see Fig. \ref{F:plaquettes}).

The main technical advantage of the $E_\text{F}$-averaging procedure is that it converts integration over the Fermi line $\Phi$ in Eq.\ \eqref{GR-av-n} into 2D integration over $\bquasi$ through the relation $dE_\text{F}=v\,d\quasi_\perp$, where $\quasi_\perp$ is the component of the quasimomentum perpendicular to $\Phi$.
Under the above conditions, the integration region is close to the ring with the radii $p_\pm=p_\text{F}\pm\Delta E_\text{F} /2v_\text{F0}$ containing approximately $(2ma^2/\pi)\Delta E_\text{F}$ plaquettes of size $\pi/a\times\pi/a$. The resulting expression for the ultrasonic attenuation rate 
$\tau_{\text{ph}\,(\nu,i)}^{-1}(\omega) = \corr{w_{\nu,i}(q)}_{E_\text{F}}$ (with $\omega=c_\nu\phon$)
reads 
\be
\label{GR-av-av}
  \tau_{\text{ph}\,(\nu,i)}^{-1}(\omega)
  =
  \frac{mc_\nu}{\pi}
  \biggl<
  \frac{|M_{\nu,i}(\phon\bn(\bquasi),\bquasi)|^2}
  {\sqrt{v^2(\bquasi)-c_\nu^2}}
  \biggr>_{\Box,E_\text{F}} ,
\ee
where the mean includes integration over a plaquette,
\be
\label{averege-over-plaqett}
  \corr{F(\bquasi)}_\Box
  =
\sqint
\frac{d^2\quasi}{(\pi/a)^2}
F(\bquasi),
\ee
with subsequent averaging over many plaquettes near the Fermi level.

\section{Dependence on the barrier transparency}
\label{S:vs-kappa}

In general, the low-temperature average relaxation rate of a phonon with polarization $\nu$ due to the interaction ${\cal V}_i$ given by Eq.\ \eqref{GR-av-av} can be represented as
\be
\label{res-f}
  \tau_{\text{ph}\,(\nu,i)}^{-1}(\omega) 
  =
  \lambda 
  \frac{c_\text{\sc{l}}}{v_\text{F0}} 
  \frac{c_\text{\sc{l}}}{c_\nu} \, \omega \,
  f_{\nu,i}(\varkappa/p_\text{F}) ,
\ee
where $v_\text{F0}=p_\text{F}/m$ and we introduced a parameter
\be
  \lambda 
  = 
  \frac{mp_\text{F}^2}
  {8\pi\rho}
  \left(\frac{v_\text{F0}}{c_\text{\sc{l}}}\right)^2
  \sim
  \frac{m}{M_\text{ion}}
  \left(\frac{v_\text{F0}}{c_\text{\sc{l}}}\right)^2
  \sim
  1 ,
\ee
with the last estimate following from the Bohm-Staver relation \cite{Ziman}. 

The behavior of the dimensionless functions $f_{\nu,i}(\varkappa/p_\text{F})$ is discussed below.

\subsection{Low barriers}

In the absence of barriers ($\varkappa=0$), there is only the Fr\"ohlich interaction with longitudinal phonons, leading to a standard relaxation rate in the 2D geometry:
\be
\label{res-clean}
  \tau_{\text{ph}\,(\text{\sc{l}},1)}^{-1}(\omega)
  =
  \lambda \zeta^2 \frac{c_\text{\sc{l}}}{v_\text{F0}} \, \omega
  .
\ee
Written in the form of Eq.\ \eqref{res-clean}, the 2D relaxation rate is analogous to its three-dimensional (3D) counterpart \cite{Migdal}: Both are linear in $\omega$ and small in the ratio $c/v_\text{F0}$ that makes phonons well-defined quasiparticles.
In terms of the function $f_{\text{\sc{l}},1}(\varkappa/p_\text{F})$ introduced in Eq.\ \eqref{res-f}, the above result means $f_{\text{\sc{l}},1}(0) = \zeta^2$.

Quite counterintuitively, scattering on vibrations of grain boundaries arises in the first---rather than second---order in the barrier height. This is a consequence of a strong modification of the 1D electron dispersion, which in the limit $\varkappa\ll p_\text{F}$ can be approximated by
\be 
\label{v-low}
  v(k) \approx 
  \frac{k}{m}
  \frac{|\delta k|}{\sqrt{\delta k^2+\alpha^2}} ,
\ee
where $\delta k=k-\pi n/a$ is a distance to the nearest edge of the Brillouin zone ($n\gg1$), and $\alpha=\varkappa/ka\ll \pi/a$. Similarly, $\beta(k)\approx(\varkappa/a^2)/(\delta k^2+\alpha^2)$.
Velocity suppression at the edges of the Brillouin zones is an interference effect due to the ideal periodicity of the considered structure.
Those narrow edges of width $\alpha\propto\varkappa$ make the main contribution to plaquette averaging in Eq.\ \eqref{GR-av-av} in the case of interaction with vibrations of grain boundaries ($i=2$), leading to nonanalytic behavior on $\varkappa^2$.

After some algebra we obtain for the leading asymptotics
at $x\ll1$:
\be
\label{gamma-res-small}
  f_{\text{\sc{l}},1} = \zeta^2 ,
\quad
  f_{\text{\sc{l}},2}  = 3x/2,
\quad
  f_{\text{\sc{t}},2} = x/2.
\ee

\subsection{Tunneling limit}
\label{SS:TunLim}

The limit of high intergrain barriers, $\varkappa\gg p_\text{F}$, can also be treated analytically. Here, the solution of Eq.\ \eqref{q-p relation} describing a 1D motion in the Dirac-comb potential has the form
$pa \approx 
  \pi \lceil \quasi a/\pi\rceil 
  \left\{
    1 + [(-1)^{\lceil \quasi a/\pi\rceil} \cos(\quasi a)-1]/\varkappa a
  \right\}
$,
where $\lceil\cdot\rceil$ stands for rounding up.
For the majority of plaquettes, $\quasi a\gg1$, and the leading contribution to the velocity $v=(p/m)\partial p/\partial\quasi$ comes from the derivatives of cosine:
\be
\label{v-tun}
  v(\quasi)\approx (\quasi^2/m\varkappa) \, 
  |\hspace{-0.9pt}\sin(\quasi a)|. 
\ee
The function $\beta(\quasi)$ defined in Eq.\ \eqref{beta-def} takes the form $\beta(\quasi)\approx \quasi^2/\varkappa$.

We will also assume that the barrier height is limited by the condition that the Fermi velocity is still much greater than the sound speed, $v(p_\text{F})\gg c$. That allows one to neglect $c_\nu$ in Eq.\ \eqref{ECL-curve}, making $\bn(\bk)$ orthogonal to $\bv(\bk)$, and in the denominator of Eq.\ \eqref{GR-av-av}.

When calculating the integral over a single plaquette in Eqs.\ \eqref{GR-av-av} and \eqref{averege-over-plaqett}, only the velocity $\bv(\bquasi)$ and hence $\bn(\bk)$ significantly depend on the position of $\bquasi$ within the plaquette via the factors $\sin\phi_\alpha\equiv\sin(k_\alpha a)$. Other types of  $\bk$-dependencies are slow and can be neglected, say, by taking $\bquasi$ in the center of the plaquette. Then averaging over the plaquette amounts to integrating over $\phi_x$ and $\phi_y$, while the remaining averaging over many plaquettes near $E_\text{F}$ is done by fixing $k=p_\text{F}$ and averaging over the direction $\mathbf{l}=\bquasi/\quasi=(\cos\theta,\sin\theta)$ of the vector $\bquasi$.
Taking the matrix elements from Eqs.\ \eqref{matelem2}, we obtain after some algebra
\be
  f_{\nu,i}(x)
  =
  x
  \int_0^{\pi} \frac{d\theta}{\pi}
  \int_0^{\pi} \frac{d\phi_x}{\pi}
  \int_0^{\pi} \frac{d\phi_y}{\pi}
  \Upsilon_{\nu,i}(\theta,\phi_x,\phi_y) ,
\ee
where
\begin{subequations}
\begin{gather}
  \Upsilon_{\text{\sc{l}},1}(x)
  =
  \frac{\zeta^2}
  {(l_x^4\sin^2\!\phi_x+l_y^4\sin^2\!\phi_y)^{1/2}} ,
\\
  \Upsilon_{\text{\sc{l}},2}(x)
  =
  \frac{4l_x^4 l_y^4 
    (l_x^2\sin^2\!\phi_x + l_y^2\sin^2\!\phi_y)^2}
  {(l_x^4\sin^2\!\phi_x+l_y^4\sin^2\!\phi_y)^{5/2}} ,
\\
  \Upsilon_{\text{\sc{t}},2}(x)
  =
  \frac{4l_x^4 l_y^4 (l_x^2 - l_y^2)^2 \sin^2\!\phi_x\sin^2\!\phi_y}
  {(l_x^4\sin^2\!\phi_x+l_y^4\sin^2\!\phi_y)^{5/2}} .
\end{gather}
\end{subequations}
Numerically calculating the integrals, we find the leading asymptotic for $x\gg1$:
\be
\label{gamma-res}
  f_{\text{\sc{l}},1} = 3.29 \, \zeta^2 x,
\quad
  f_{\text{\sc{l}},2} = 2.00 \, x ,
\quad
  f_{\text{\sc{t}},2} = 0.47 \, x .
\ee

The low-temperature tunneling-limit results given by Eq.\ \eqref{gamma-res} have a number of pronounced features.
Firstly, the phonon relaxation rate does not depend on the size of the granules $a$. Secondly, since $\zeta\sim1$, the contribution from the displacement of grain boundaries is comparable to that due to the standard Fr\"ohlich's mechanism. Thirdly, longitudinal phonons are by the factor of four more efficient than transverse phonons for the relaxation induced by boundaries vibrations.

Equations \eqref{res-f} and \eqref{gamma-res} predict a linear growth of $\tau_\text{ph}^{-1}(\omega)$ with an increase in the barrier height $\varkappa$ 
that is a consequence of a proportional decrease of the electron velocity in the tunneling regime.
This is an intermediate asymptotics obtained under two constraints: (i) $\gamma\gg1$ [see Eq.\ \eqref{gamma-def}] and (ii) $v\gg c$. The former condition implies that despite ${\cal T}(p_\text{F})\ll1$ we are still in the regime of a good metal and the approximations used are justified. The latter condition means that electron-phonon coupling is weak, Migdal's theorem is applicable, and phonons are well-defined quasiparticles, with the attenuation rate $\tau_\text{ph}^{-1}(\omega)$ being much smaller than $\omega$.

\begin{figure}
\centering
\includegraphics[width=0.95\linewidth]{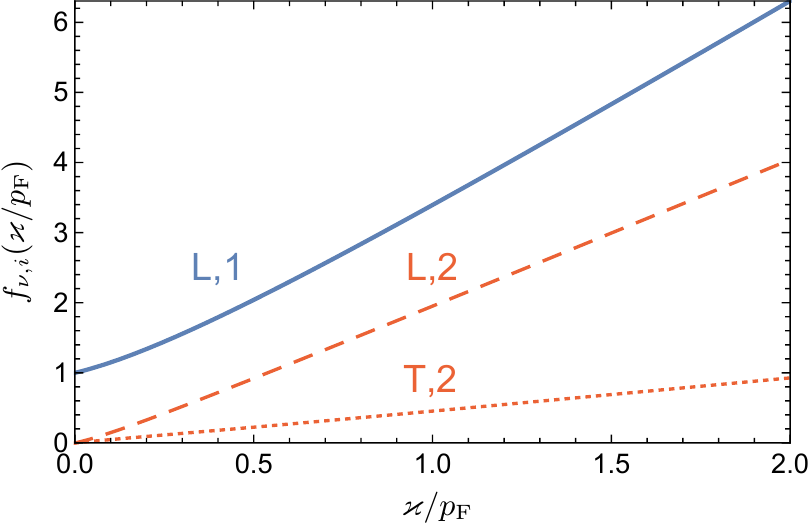}
\caption{Functions $f_{\nu,i}(\varkappa/p_\text{F})$, which describe the ultrasound attenuation rate according to Eq.\ \eqref{res-f}, vs the dimensionless barrier height $\varkappa/p_\text{F}$ (the Fr\"ohlich interaction constant $\zeta=1$).}
\label{F:num}
\end{figure}

\subsection{Arbitrary boundary transparency}

At an arbitrary ratio of $\varkappa/p_\text{F}$, the ultrasound attenuation rate given by Eq.\ \eqref{GR-av-av} should be obtained numerically. The result expressed in terms of the functions $f_{\nu,i}(\varkappa/p_\text{F})$ introduced in Eq.\ \eqref{res-f} is presented in Fig.\ \ref{F:num}. 
Simulations are performed at $p_\text{F}a=100$, and we used the RPA-value of $\zeta=1$ for the Fr\"ohlich interaction constant in 2D. The graphs demonstrate a crossover from the case of transparent boundaries [Eq.\ \eqref{gamma-res-small}] to the tunneling limit [Eq.\ \eqref{gamma-res}]. The curves {\sc l},2 and {\sc t},2 describing electron-phonon interaction due to vibration of grain boundaries look almost like straight lines that is artifact of numerical closeness of the respective coefficients in the small- and large-$x$ asymptotics.

\subsection{Effect of the electron spectrum broadening}

In our calculation of the heat transfer rate, we have neglected the effects of inelastic scattering on the electron spectrum. Such a relaxation with the rate $\tau_\text{in}^{-1}$ would smear the delta function representing the energy conservation in Fermi's golden rule \eqref{GR-CF} into the corresponding Lorentzian. As a result, the line $\Gamma_\bn$ defined by Eq.~\eqref{ECL-curve} will acquire a finite width $\delta k_t \sim (\tau_\text{in}q|\partial_\bquasi(\bn\bv)|)^{-1}$, with $q\sim T/c$ being a typical (thermal) phonon momentum. Roughly speaking, this broadening can be neglected provided it is smaller than the size of the plaquette, $\pi/a$. However, $\delta k_t$ is sensitive to the position of the point $\bk_0$, where the line $\Gamma_\bn$ and the Fermi line $\Phi$ intersect. At the same time, even if $\delta k_t$ becomes large on a part of the line $\Gamma_\bn$, which does not make a significant contribution to the electron-phonon scattering rate, that will not change our results for $\tau_\text{ph}^{-1}(\omega)$. To take this into account, we evaluate the average $\delta k_t$ weighted with $\mu(\bk) = |M(\phon\bn(\bquasi),\bquasi)|^2/v(\bquasi)$, the factor which determines the average phonon decay rate according to Eq.\ \eqref{GR-av-av} (we assume $v\gg c$). As a result, the condition of neglecting the interaction-induced broadening of the electronic spectrum is formulated as
\be
  \tau_\text{in}^{-1} 
  \ll 
  \frac{T}{ca} 
  \corr{ |\partial_\bquasi[\bn\bv(\bk)]|^{-1} }^{-1}
  ,
\ee
where the average is defined as
\be
  \corr{ X(\bk) }
  = 
  \frac{\corr{ X(\bk) \mu(\bk) }_{\Box,E_\text{F}}}
  {\corr{ \mu(\bk) }_{\Box,E_\text{F}}} .
\ee
In the limit of vanishing $c$, the vectors $\bn$ and $\bv$ are orthogonal at the point $\bk_0$. Then calculating the derivative and taking it at $\bk=\bk_0$, we get
\be
\label{int-cond}
  |\partial_\bquasi[\bn\bv(\bk)]|
  =
  v^{-1} \!
  \sqrt{v_y^2(\partial v_x/\partial k_x)^2 + v_x^2(\partial v_y/\partial k_y)^2} .
\ee

To evaluate the effect of inelastic broadening, we focus on the Fr\"ohlich interaction with longitudinal phonons. 
In the tunneling limit, $\varkappa\gg p_\text{F}$, the velocity is given by Eq.\ \eqref{v-tun} and the average $\corr{ |\partial_\bquasi[\bn\bv(\bk)]|^{-1} }$ is determined by the whole plaquette. Then a simple estimate yields
$  
  \corr{ |\partial_\bquasi[\bn\bv(\bk)]|^{-1} }
  \sim
  m \varkappa/p_\text{F}^2 a
$.
In the low-barrier limit, $\varkappa\ll p_\text{F}$, a plaquette is organized into a central part with the free velocity $\bv=\bk/m$ and a small region of width $\alpha=\varkappa/p_\text{F}a\ll \pi/a$ near the edges of the Brillouin zones, where the velocity is suppressed due to interference, see Eq.\ \eqref{v-low}. In the latter region, the derivative $|\partial_\bquasi[\bn\bv(\bk)]|$ is large and its contribution to the average $\corr{|\partial_\bquasi[\bn\bv(\bk)]|^{-1}}$ can be shown to be negligible. Hence the average is determined by the central part of the plaquette, leading to 
$  
  \corr{ |\partial_\bquasi[\bn\bv(\bk)]|^{-1} }
  \sim
  m
$.
With increasing $\varkappa$, the area of the central part shrinks and it disappears at $\varkappa\sim p_\text{F}$.
Summarizing the above results, we may write the following rough interpolation formula:
\be  
  \corr{ |\partial_\bquasi[\bn\bv(\bk)]|^{-1} }
  \sim
  m ( e^{-x} + x/p_\text{F} a ) 
\ee
qualitatively valid for all $x=\varkappa/p_\text{F}$.

Assuming electron-phonon scattering as the main source of electron relaxation and using Eq.\ \eqref{Sigma}, we get $\tau_\text{in}^{-1}\sim (\tau_\text{ph}^{-1}/\omega)T^2/mc^2$.
Employing now Eq.\ \eqref{res-f}, we see that the inequality \eqref{int-cond} is well satisfied in the low-temperature limit considered [see Eq.\ \eqref{low-T}] both for low and high barriers.
In the limit of $\varkappa/p_\text{F}\ll1$, it translates into $Ta/v_\text{F}\ll1$ guaranteed by $c\ll v_\text{F0}$. In the limit of $\varkappa/p_\text{F}\gg1$, it imposes the constraint 
$Tx^2/p_\text{F}v_\text{F0}\ll1$ satisfied since $v\gg c$ and $\gamma\gg1$ [see Sec.\ \ref{SS:TunLim}].
Electron-electron interaction with the rate $\tau_\text{ee}^{-1}\sim T^2/E_\text{F}$ in the clean limit is even less important.

Hence, inelastic broadening of the electron spectrum does not modify our results for the electron-phonon relaxation rate in the low-temperature limit [Eq.\ \eqref{low-T}].

\section{Conclusion}
\label{S:Conclusion}

We have studied the electron-phonon interaction in a model of a regular granular medium, with the granules formed by periodically placed $\delta$-functional potential barriers. The problem is solved in the two-dimensional geometry under the assumption of ballistic electron motion inside the granules. An essential element of the theory is the nonperturbative account of potential barriers, with a considerable modification of the electronic states compared to plane waves in the clean case. Phonons are considered in the standard Debye model, assuming a continuous description in terms of acoustic longitudinal and transverse modes. Both the usual Fr\"ohlich's electron-phonon interaction and the interaction caused by the displacement of grain boundaries dragged by the lattice are taken into account.
The proposed model is an oversimplified description of real disordered granular structures, where neither the momentum nor the quasi-momentum is a good quantum number. 

The main result of the work is Eq.\ \eqref{res-f} with the obtained functions $f_{\nu,i}(\varkappa/p_\text{F})$ describing the ultrasound attenuation rate in the low-temperature limit, $T\ll c/a$. It has a pretty universal form independent of the period of the granular structure:
\be
\label{res-simp}
  \tau_\text{ph}^{-1}(\omega) \sim (c/v) \omega ,
\ee
where $v$ is the typical electron velocity at the Fermi surface determined by the barriers transparency: $v\sim v_\text{F0}=p_\text{F}/m$ for $\varkappa\ll p_\text{F}$, and $v\sim (p_\text{F}/\varkappa) v_\text{F0} \sim E_\text{F}/\varkappa\ll v_\text{F0}$ for $\varkappa\gg p_\text{F}$.
The main qualitative effect is the growth of $\tau_\text{ph}^{-1}(\omega)$ with the decrease in the intergrain transparency, which is a consequence of the slowing down of electron transport between granules in the tunneling limit.

Independence of Eq.\ \eqref{res-simp} on the granule size $a$ suggests that it also holds for a similar three-dimensional model of regularly arranged granules.
Then the linear frequency dependence $\tau_\text{ph}^{-1}(\omega)$ results in the heat transfer rate \eqref{T^p-T^p} with the usual clean exponent $p=d+2$. Such a behavior, with the prefactor exceeding the clean-limit value, qualitatively agrees with the experimental results reported for polycrystalline NbN films \cite{Lomakin}.

Our analysis is applicable in an experimentally relevant regime of a good metal, when the DOS, though exhibiting a number of van Hove singularities, is nearly constant around $E_\text{F}$. This situation is realized provided $(p_\text{F}/\varkappa)(p_\text{F}a)\gg1$. Under this condition it is possible to perform energy averaging and get rid of a quasi-regular oscillating energy dependence. The same condition justifies the use of the standard Fr\"ohlich Hamiltonian obtained for a constant DOS.

In our calculation of the heat transfer rate, we have neglected the effects of inelastic scattering on the electron spectrum. This approximation is justified within the range of applicability of the developed theory.

The peculiarity of the considered model is the periodicity of the grain structure, which leads to the quasimomentum conservation and the absence of diffusion. In real materials, grain boundaries are placed irregularly, resulting, even for ballistic motion inside the granules, in diffusive electron propagation on large scales. The description of the electron-phonon interaction in disordered granular metals remains a challenging open question requiring further study.

\acknowledgments

The authors are grateful to A. V. Semenov for discussing the problem statement. This research was supported by the Russian Science Foundation under Grant No. 23-12-00297.

\end{document}